\pdfoutput=1
\documentclass[runningheads]{llncs}
\usepackage[T1]{fontenc}
\usepackage{graphicx}
\usepackage{amsmath}

\usepackage{caption}
\usepackage{subcaption}

\usepackage{booktabs}
\usepackage{tabularx}
\usepackage{makecell}
\usepackage{multirow}
\usepackage{comment}
\usepackage[hyphens]{url}
\usepackage{hyperref}
\usepackage{cleveref}
\usepackage{mathabx}

\usepackage{epigraph}
\hypersetup{
  colorlinks= true,
  linkcolor = blue,
  urlcolor  = blue,
  citecolor = black,
}

\newcommand*{\fullref}[1]{\hyperref[{#1}]{\cref*{#1} \nameref*{#1}}}
\newcommand*{\Fullref}[1]{\hyperref[{#1}]{\Cref*{#1} \nameref*{#1}}}
\newcommand*{\secref}[1]{\hyperref[{#1}]{\autoref*{#1}}}
\newcommand*{\Secref}[1]{\hyperref[{#1}]{\Cref*{#1}}}

\graphicspath{{images/}}

\usepackage{color}

\begin{document}
\title{Speaker Mining - FAIR Data on\\ Public Broadcasts for Question Answering}
%
%

\author{Tim Wittenborg\inst{1}\orcidID{0009-0000-9933-8922} \and Omar Imad Remmo \and Claudia Frick\inst{3}\orcidID{0000-0002-5291-4301} \and 
Lena John\inst{2}\orcidID{0009-0007-2097-9761} \and 
Oliver Karras\inst{2}\orcidID{0000-0001-5336-6899} \and
Sören Auer\inst{1,2}\orcidID{0000-0002-0698-2864}}
\authorrunning{T. Wittenborg et al.}
%
\institute{L3S Research Center, Leibniz University Hannover, Hannover, Germany\\
\email{tim.wittenborg@l3s.uni-hannover.de} \and
TIB - Leibniz Information Centre for Science and Technology, Hannover, Germany\\
\email{\{lena.john, oliver.karras, soeren.auer\}@tib.eu} \and
Institute for Information Science, TH Köln - University of Applied Sciences Cologne, Cologne, Germany\\
\email{claudia.frick@th-koeln.de}}
\maketitle  
\begin{abstract}
Public broadcasts are at the center of civic discourse: 
Traditional television talk shows, alongside emerging podcast and web video formats, capture and guide the attention of our societies, shaping how citizens encounter politics, science, and societal issues. 
Yet, systematic or even simple analyses of these formats face similar challenges: guest and content metadata are scarce, fleeting, fragmented, and not standardized.
Research conducted and questions answered are based on extensive, laborious, yet isolated data-curation efforts that capture only a fraction of the relevant landscape.
This work seeks to address this issue using a scaling-oriented framework for FAIR data curation in public broadcasting.
Evaluated on 15 broadcasting programs, the pipeline aggregates ZDF Archive PDFs, fernsehserien.de, and Wikidata into a unified knowledge graph.
Of the 31,817 candidate guest mentions from these three sources, 17,729 could be automatically disambiguated, further 5,958 via 64 hours of manual reconciling using OpenRefine.
Results are published at speakermining.wikibase.cloud and linked to Wikidata, enabling SPARQL-based question answering based on gender, age, occupation, or institutional affiliation across 8,436 canonical persons with 23,527 appearances in 6,469 aligned episodes.
Our iterative experience reveals that correctly disambiguating and deduplicating speaker data from heterogeneous sources demands dedicated effort on sustainable infrastructure.
For scalable and reliable question answering on public broadcasts to be accessible to everyone, we recommend fostering the potential of linked open data:
Advancing alignment and utilization approaches like this work, particularly towards crowdsourced development and curation, 
but also more FAIR data interfaces from public broadcast service providers.
\end{abstract}

\section{Introduction\label{sec:introduction}}
Public broadcast talk shows and interview formats shape public discourse. 
Programs such as political talk shows and science podcasts are major drivers of how citizens encounter politics, science, and societal issues~\cite{Phogat.2025}. 
Millions of viewers encounter experts and scientific perspectives through these formats rather than through direct engagement with scientific publications~\cite{Rosen.2016}.
This underlines the democratic relevance of public broadcasts and science journalism as a specific form of science communication~\cite{Davidson.2024}.

Systematic analysis of these broadcast formats remains difficult because relevant metadata is fragmented, inconsistent, and often inaccessible.
But these formats are especially important because editorial selection determines which experts, topics, and perspectives become publicly visible, making them prone to biases and stereotypical representations of science. 
Journalistic gatekeeping can privilege certain demographics, institutions, and viewpoints while limiting the visibility of others~\cite{Riedl.2024,Rosen.2016}.
To investigate these effects, which might stem from any of the identity types evident in the academic wheel of privilege~\cite{elsherifBridgingNeurodiversityOpen2022}, studying science communication across journalistic formats is necessary.
Broadcast metadata is rarely FAIR (Findable, Accessible, Interoperable, Reusable).
Researchers, therefore, spend substantial effort manually collecting and reconciling guest information before analysis can begin. 
Structured FAIR broadcast data would also enable question answering (QA) about guest variability, age distribution, or topic visibility at scale~\cite{WissKommProposal.2026,Wittenborg.2025}.

This work presents a framework for FAIR curation of public broadcast guest data and accounts for where true scalability remains unsolved.
Using 15 broadcast programs as a case study, we aggregate and reconcile heterogeneous metadata from ZDF Archive PDFs, fernsehserien.de, and Wikidata into a shared Wikibase infrastructure.
Our contributions include: 
(1) An open source pipeline\footnote{\url{https://github.com/borgnetzwerk/speaker-mining}} for event-sourced candidate generation, alignment across sources, and staged entity disambiguation and deduplication.
(2) A curated FAIR dataset\footnote{\url{https://speakermining.wikibase.cloud/}} covering 6469 episodes across 12 broadcasting programs (only 10 of which have a total of 4,861 episodes with guest data), disambiguated using OpenRefine.
(3) Quantitative analyses across the 4,861 episodes for 15 selected guest properties, particularly gender, age, occupation, position, and party affiliation.
The remainder is structured as follows: \autoref{sec:background_rw} presents background and related work, \autoref{sec:approach} outlines our approach, \autoref{sec:implementation} describes the implementation, \autoref{sec:results} presents results, \autoref{sec:discussion_futurework} discusses findings and future directions, \autoref{sec:conclusion} concludes.

\section{Background and Related Work\label{sec:background_rw}}
\textbf{Science communication} is increasingly studied as a socio-technical system shaped by diverse actors, media formats, and institutional constraints. 
Prior work highlights that different communication channels, including journalism, social media, visual communication, participatory formats, and humour, come with distinct affordances and challenges for communicating scientific knowledge~\cite{enzingmullerVisuelleWissenschaftskommunikationForschungsuberblick2025,Phogat.2025,rufHumorUndWissenschaftskommunikation2025}.
Within this landscape, journalism occupies a particularly influential role because it mediates between scientific communities and broader publics, thereby shaping which topics, experts, and perspectives receive visibility~\cite{Brueggemann.2020,schwind2024wisskomm}. 
Related work documents persistent inequalities: women are underrepresented as experts in journalism, and media structures can reproduce existing systemic biases, including gender and ethnic disparities in quotation practices and media visibility~\cite{Arabi.2025,Corsbie.2022,Davidson.2024,Peng.2024}.
However, many such studies rely on purpose-built datasets that remain difficult to reproduce, extend, or reuse across formats and time periods.
Public broadcast talk shows occupy a distinctive position within this landscape: they operate under editorial curation, reach millions of viewers per episode, and create a recurring, timestamped record of which voices receive public visibility over years and decades.
Yet systematic longitudinal analysis of these records has been hampered by the absence of structured, reusable metadata.

\textbf{FAIR data practices}, particularly Knowledge Graphs (KGs), are promising approaches for addressing these challenges.
Public broadcaster archives, web aggregators such as \href{https://www.fernsehserien.de/}{fernsehserien.de}, and open knowledge bases such as \href{https://www.wikidata.org/}{Wikidata} offer complementary but heterogeneous metadata sources.
Our prior work on the emerging Science Communication Knowledge Infrastructure (SciCom KI) argued that existing curation infrastructures remain fragmented and insufficiently scalable for audiovisual formats~\cite{Wittenborg.2025}.
The \href{https://blog.rufus-portal.de/}{rufus}~\cite{Blume.2024} portal addresses similar challenges, supported by the ZDF archive providing data on almost 500.000 episodes.

\textbf{Related analyses} demonstrate the societal relevance and analytical potential of structured broadcast metadata.
\href{www.spiegel.de}{Der Spiegel}~\cite{spiegelTalkshowAnalysis} repeatedly investigated recurring guests and representational imbalances in German talk shows.
Independent projects such as \href{https://lanz-mining.arrrrrmin.dev}{LanzMining}~\cite{lanzmining2025} explored automated aggregation of talk show guests using large language models and simple regex-based classification.
In collaboration with LanzMining, a community-driven approach coordinated by Stefan ``stk'' Kaufmann~\cite{stk} demonstrated the use of Wikidata and linked open data for analyzing guest appearances and party representation in political talk shows.
\autoref{tab:prior_comparison} contrasts our framework with the prior analyses whose datasets permit cross-validation.
Only one other approach provides Wikidata-linked, SPARQL-queryable output. 
To our knowledge, no other contained an explicit entity deduplication protocol.
Our framework extends coverage to 15 programs (12 with episode data, 10 of those with guest data) and links 62.3\% of unique persons to Wikidata.
Our work builds on this prior work towards a tool-supported, rule-based disambiguation workflow for FAIR broadcast data.

\begin{table}[bth]
 \vspace{-0.5cm}
\centering
\caption{Comparison of related talk show guest analysis approaches.}
\label{tab:prior_comparison}
\begin{tabularx}{\linewidth}{l X X l X l l}
\toprule
 & \textbf{Period} & \textbf{Shows} & \textbf{Episodes} & \textbf{Guests}  & \textbf{Wikidata} & \textbf{Output} \\
\midrule
Spiegel~\cite{spiegelTalkshowAnalysis} & 2015--25 & 6 & 2854 & 3,991 & --- & CSV \\
LanzMining~\cite{lanzmining2025} & 2022--25 & 5 & 515 & 813  & --- & CSV \\
stk~\cite{stk} & 2024 & 1 & 138 & 262 & 100\% & Wikidata \\
\hline
Our work & 1972--26 & 15 & 6,469 & 8,287 & 62.3\% & Wikibase \\
$\drsh$ with guest data & 2006--26 & 10 & 4,861 & 8,287 & 62.3\% & Wikibase \\
\bottomrule
\end{tabularx}
\end{table}

\section{Approach}\label{sec:approach}
We pursued a digital library approach to FAIR broadcast data in two iterations: (1) an explorative single-source run to identify roadblocks, and (2) a full redesign with multiple sources and a contract-coupled, event-sourced pipeline.

\subsection{First Iteration: Explorative Discovery}
Detailed in a bachelor's thesis~\cite{Omar.2026}, four PDFs from the ZDF archive were processed in a four-stage pipeline (PDF extraction, Wikibase setup, OpenRefine entity linking, SPARQL analysis).
It extracted 2,035 of 2,036 episodes and linked 75.3\% to Wikidata through eight hours of manual review.
The 36\% female guest share matched Spiegel's independent figure of 35\%, validating the approach. \autoref{tab:iter1_results} summarizes the key quantitative results.
\begin{table}[h]
 \vspace{-0.5cm}
\centering
\caption{Selected quantitative results of the first iteration.}
\label{tab:iter1_results}
\begin{tabular}{lr}
\toprule
\textbf{Metric} & \textbf{Value} \\
\midrule
Episodes extracted & 2,035 of 2,036 \\
Person items created & 3,966 \\
Persons linked to Wikidata & 2,984 (75.3\%) \\
Episodes with no guests detected & 125 (6.1\%) \\
Female guest share (appearances / unique)\hspace{0.2cm} & 36.0\% / 31.3\% \\
\bottomrule
\end{tabular}
 \vspace{-0.5cm}
\end{table}

The iteration equally revealed systematic roadblocks that motivated the complete redesign:

\begin{enumerate}
    \item \textbf{Brittle extraction.} The regex parser was tightly coupled to the ZDF archive typographic conventions; minor formatting variations caused 125 episodes (6.1\%) to yield no guests; multi-part surnames (e.g., \textit{von Schönburg}) were systematically malformed or not even registered if they contained non-capital ASCII letters (e.g., \texttt{ß}, \texttt{Ö}).
    \item \textbf{Missing pre-import disambiguation.} Person identity resolution was deferred to OpenRefine; near-duplicate name typos remained as separate items, and distinct persons sharing a name were automatically merged by default.
    \item \textbf{Manual entity linking does not scale.} With 40.4\% of persons requiring manual review, the eight-hour effort is impractical for larger corpora or periodic updates.
    \item \textbf{Absent temporal semantics.} Wikidata party memberships were queried without temporal filtering, causing historically inaccurate attributions: Bündnis Sahra Wagenknecht (founded in 2024) appeared in the 2012 timeline.
    \item \textbf{Single-source design.} The pipeline was built around the ZDF archive structure and did not integrate episode data from fernsehserien.de or Wikidata into the Knowledge Graph.
\end{enumerate}

\subsection{Second Iteration: Informed Redesign}
\begin{figure}[bth]
     \vspace{-0.5cm}
    \centering
    \includegraphics[width=\linewidth]{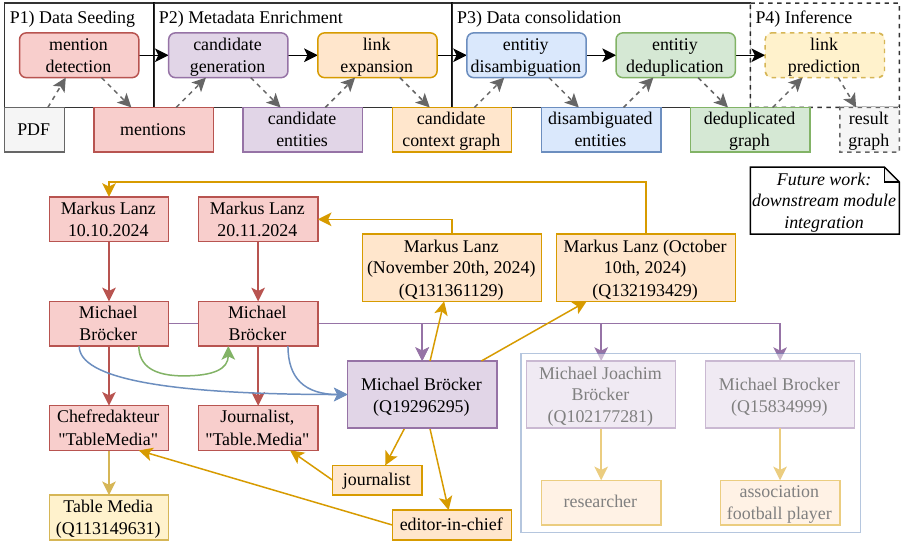}
    \caption{Final approach after redesign.}
    \label{fig:final_approach}
     \vspace{-0.5cm}
\end{figure}

Taking the lessons learned from the first iteration into account, the approach was reworked from the ground up to the shape illustrated in \secref{fig:final_approach}.

It is composed of primarily three coupled phases with downstream potential:
\begin{itemize}
    \item \textbf{Phase 1} extracts and structures mentions from the ZDF archive export: Episodes, their publications, guests, topics, and related organizations.
    \item \textbf{Phase 2} generates relevant candidates for these instances in different sources and explores their connections to related items, potentially identifying additional relevant instances.
    \item \textbf{Phase 3} synthesizes this information by disambiguating matches describing the same event across sources (person X, guest of episode 1) and deduplicating within the resulting instance list (person X, guest of episode 1, equals person Y, guest of episode 3).
    \item \textbf{Phase 4} is scoped as future work: it would inspect the remaining instance set for unresolved links and properties, attempting to predict additional linkages beyond the disambiguation and deduplication of Phase~3.
\end{itemize}

The implementation section describes these phases in detail.

\section{Implementation}\label{sec:implementation}

The pipeline is implemented as a sequence of Jupyter notebooks, each orchestrating a set of dedicated Python modules.
Intermediate output projections are written to CSV and JSON files under a phase-bound directory contract, so any notebook can be rerun independently once its upstream inputs are available.

\subsection{Phase 1: Text Extraction and Mention Detection}

PDF-to-text conversion uses \texttt{pdfplumber}; each episode receives a SHA1-based stable identifier derived from title and first broadcast date.
Guest extraction follows a four-strategy rule hierarchy, from a structured info-section parser to a surname fallback for ALL-CAPS tokens, in descending confidence order.
A pattern-based institution extraction module was also initially developed; however, due to a high false-positive rate, this module was deferred until after semantic validation via Wikidata was concluded for additional context.

\subsection{Phase 2: Candidate Generation}

Wikidata candidate generation performs a relevancy-seeded BFS from the configured broadcasting programs:
episodes inherit relevancy from their series, guests from relevant episodes. 
Nodes are classified \texttt{basic\_fetched} (label, description, alias, and class hierarchy), \texttt{all\_outlinks\_fetched}, or \texttt{full\_fetched} (in- and outlinks) based on relevance rules.
A precision-tuned threshold prevents runoff into unrelated entities; a permissive threshold causes relevance to spread to thousands of unrelated entities, while a strict threshold drops valid guests. 
Our implemented design leans toward precision: only seed broadcasts are \texttt{full\_fetched}, shows listing them as \texttt{part of series} are \texttt{all\_outlinks\_fetched}, as well as their guests and a selection of guest property objects (e.g. \textit{occupation}, \textit{position}).

Wikidata graph expansion can take hours, and interruptions risk discarding current or corrupting prior progress. 
Logging infrastructure, issue tracing, checkpoints, iterative improvement, and append-only caching were needed.
Initial runs required over 25,000 queries over several hours, creating an equal number of JSON artifacts to avoid data loss during a write interruption.
While event-sourcing was not the initial design choice, it naturally emerged as a favorable solution and met all of the identified requirements.
Every query result, entity fetching, triple, relevancy assignment, and rate-limit observation is written as an immutable event. 
The pipeline is now built around an append-only JSONL event log.
Event handlers can replay this log at any time to regenerate all projection artifacts.
Subsequent runs replay the cache in seconds.

The fernsehserien.de pipeline reused these event-sourcing and cache-first patterns from Wikidata.
Once the Wikidata pipeline structure was robust and iteratively improved, adding this second source required minimal additional effort.

\subsection{Phase 3: Entity Disambiguation}

Phase~3 comprises three sequential steps: Step~3.1.1: Alignment, Step~3.1.2: Reconciliation, and Step~3.2: Deduplication.

\textbf{Step~3.1.1: Alignment} normalizes and aligns entities across all sources within seven classes: programs, seasons, episodes, persons, topics, roles, organizations.
Primary alignment targets are episodes.
Normalization applies a schema-mapping table that spans all properties across all sources, as well as accent removal, lowercasing, and whitespace collapsing before comparison.
Normalized label matches and publication time matches, along with belonging to the same configured broadcasting program, were the highest-quality alignment signals.
High-confidence near-matches were technically also possible, and the remainder were flagged \texttt{UNRESOLVED} with typed reason codes (\textit{no\_candidate}, \textit{low\_confidence}, \textit{contradiction}, \textit{insufficient\_context}).
Quality gates at the end of Step~3.1.1 validate row counts and schema completeness before the handoff to manual disambiguation.

\textbf{Step~3.1.2: Reconciliation} imports these rows to OpenRefine for manual reconciliation.
Two curators independently reviewed separate alphabetical slices (A--L: 16,772 rows; M--Z: 15,043 rows).

\textbf{Step~3.2 Deduplication} clustered canonical persons by shared Wikidata QID or normalized name similarity as a lower-confidence fallback.
Other entity classes could be sufficiently disambiguated via data-driven alignment (episodes and broadcasts) or were deferred (institutions and topics).

\section{Results}\label{sec:results}
\epigraph{Results! Why, man, I have gotten a lot of results! I know several thousand things that won’t work.}{\textsc{Thomas Edison (Q8743)}}

The main result of our two-iteration approach is that additional iterations are needed. 
After resolving the known issues in our first iteration, we found a seemingly exponential cascade of issues with each one we solved.
Yet,
we eventually found that the scaling was not truly exponential, just frontloaded. 
Like subclass expansion, only the first few layers beneath a top class provided a steep increase in numbers. 
The deeper we walked down any issue path, the more we could close issues and eventually refine systems (such as our cached Wikidata interface), thereby catching most future cases with minimal overhead.
As such, we now provide insights into our use-case results and highlight the framework that aggregated them in the discussion section. 

Of the 2,036 episodes extracted from the 2008 to 2024 ZDF corpus, only thirteen episodes produced no guest rows, down from 125 in the first iteration.
Many of the remaining episodes without guests are correctly identified as moderator-only special episodes.
When extracting the 6,459 episodes from fernsehserien.de, 1,639 were without guests, 2,434 were without a moderator.
Where those could be matched to the ZDF Archive, these should have had guests, hinting at gaps in the database.
The Wikidata BFS also identified 638 series instances, with 382,400 Wikidata triples spanning 8,221 unique classes and 2,324 properties. 

In Step~3.1.1: Alignment, 2,673 of 6,849 episodes (39\%) were successfully aligned across sources.
Reduced to only episodes with guest data, we arrived at 4,861 episodes.
The automated person alignment matched 17.2\% (5,487) of the 31,817 person appearance rows as high-confidence matches.
This context was presented to the OpenRefine curators, who conducted Step~3.1.2 Reconciliation: 
First, all 31,817 person rows were automatically reconciled (55.91\%).
Over 64 hours, these matches and reconciliations were confirmed, and another 18.79\% were manually mapped.
For another 12.17\%, no suitable candidate could be found, which mostly came down to encoding errors, single-word names, or organizations mislabeled as persons.
13.13\% remain unresolved within the available time window.
We arrive at 26,659 QID-matched entries of 8,436 canonical persons, of which 5,257 (62.3\%) could be linked to Wikidata.
Reduced to only guests of our target episodes, we arrive at 23,527 appearances of 8,287 unique guests.

We classify this data along four data quality tiers: Tier~1: 2,394 (28.4\%) persons have a Wikidata QID and are annotated in at least two sources; Tier~2: 2,863 (33.9\%) have a Wikidata QID and only one source; Tier~3: 628 (7.4\%) are disambiguated via two non-Wikidata sources but have no QID; Tier~4: 2,551 (30.2\%) are single-source only, with no QID.

Alphabetically sorting and stacking same-name entries alongside their appearance timelines proved highly effective for manual review.
Data quality issues illustrate the value of structured data: We found institutions and editorial staff tagged as guests, prefixes (e.g., ``Prof.'') or relationship descriptors captured as names (e.g., ``her son Bob''), or compound descriptors (e.g., ``family Miller'').
Due to alphabetical sorting, most of these could quickly be batch-resolved.

\subsection{Analysis}
\begin{figure}[bth]
    \centering
    \caption{Per-show episode and guest statistics across the 12 analyzed programs.}
    \vspace{-0.4cm}
    \includegraphics[width=\linewidth]{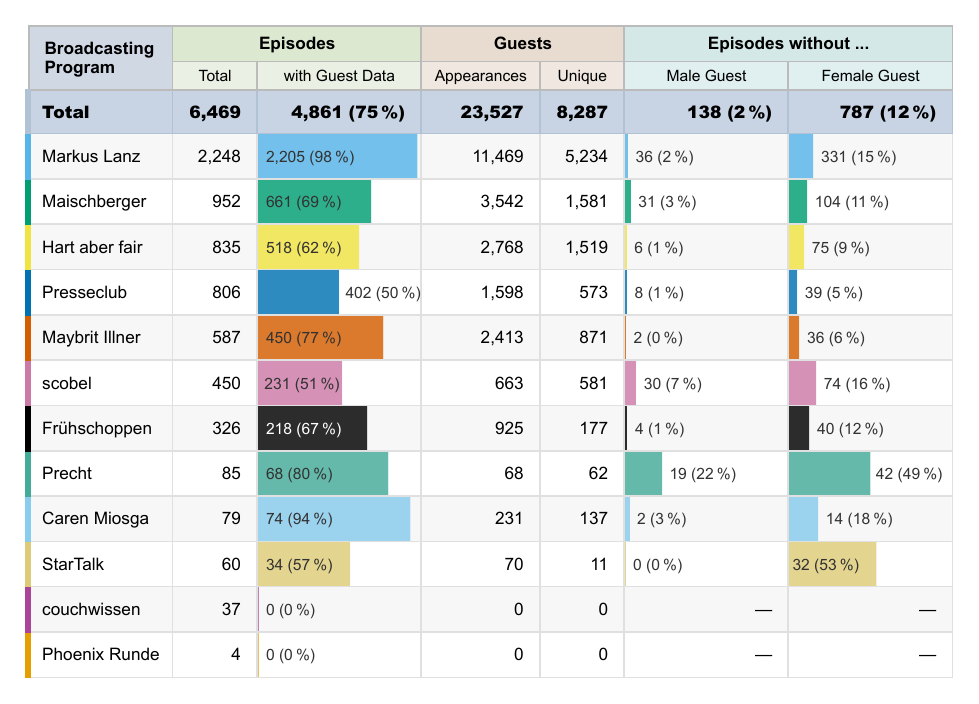}
    \label{fig:00_show_stats_table_en}
    \vspace{-0.6cm}
\end{figure}

An episode-guest occurrence matrix was built and used to navigate the analysis.
Every guest's Wikidata entity was fully resolved with all outgoing links, enabling by-property analysis across 15 specified properties.
Of the 8,436 deduplicated persons, 39 are moderators, 22 staff members, and 88 additional persons incidentally captured (such as episode topics or relatives of guests), yielding $8{,}436 - 39 - 22 - 88 = 8{,}287$ guests.
The 15 programs were selected in four groups:
The first group comprises the German political talk shows most extensively studied by prior work~\cite{spiegelTalkshowAnalysis,lanzmining2025,stk}, enabling cross-validation against independently derived figures. The shows are
\textit{Markus Lanz}, \textit{Maischberger}, \textit{Hart aber fair}, \textit{Maybrit Illner}, \textit{Caren Miosga}, and \textit{Phoenix Runde}.
The second group of \textit{Presseclub} and \textit{Der internationale Frühschoppen} follows the same format and audience profile but was absent from those analyses, making it a natural extension.
The third group of \textit{Precht}, \textit{Lanz und Precht}, and \textit{unbouble} shares structural properties with political talk shows while differing in scope or medium, airing primarily online, and are testing the pipeline generalizability across format variation.
\textit{StarTalk}, \textit{scobel}, \textit{couchwissen}, and \textit{wtf\_talk} are included to extend this corpus towards science communication, broadening the scope from the political-talk-show focus of prior analyses~\cite{lanzmining2025,spiegelTalkshowAnalysis} toward the cross-format vision of SciCom Wiki~\cite{WissKommProposal.2026}.
Five programs ultimately lacked usable guest data: \textit{unbouble}, \textit{wtf\_talk}, and \textit{Lanz und Precht} had no extractable data at all; \textit{Phoenix Runde} and \textit{couchwissen} had episode data but no guest entries.
\secref{fig:00_show_stats_table_en} shows the per-show breakdown for the 12 with at least some episode data.
Analyses of Wikidata-sourced properties (gender, age, party, occupation) apply only to the 5,257 Tier~1 and Tier~2 persons with Wikidata links; the remaining 37.7\% of canonical persons carry no property data and are excluded from all property-based analyses below.

\subsubsection{Gender and Age.}
\begin{figure}[bth]
    \vspace{-0.5cm}
    \centering
    \includegraphics[width=1\linewidth,trim={0 1cm 0 0},clip]{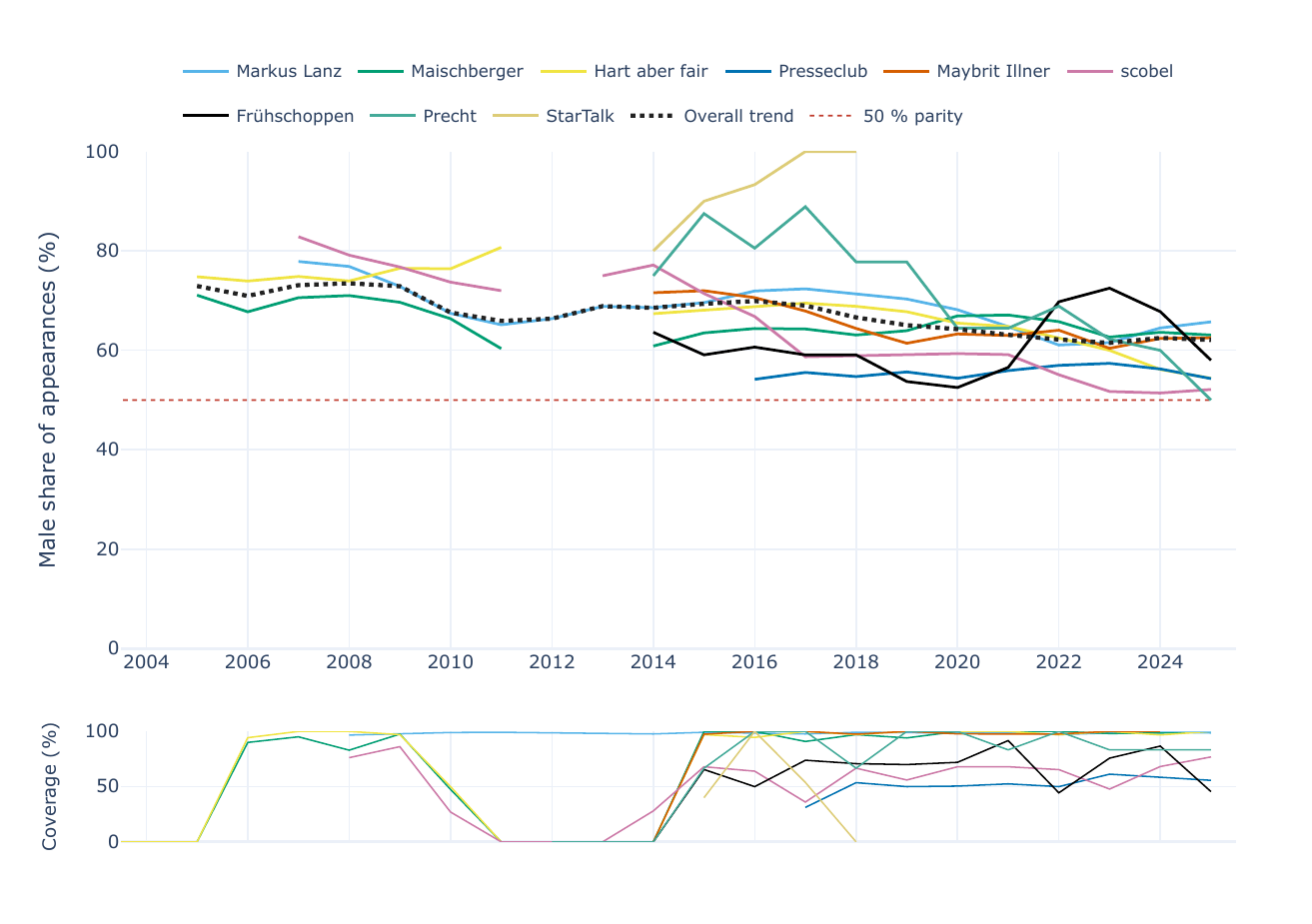}
    \caption{Gender share of guest appearances over time. The 2011 gap reflects missing source coverage on guest data for those years. The long-term trend toward gender parity appears to have ceased around 2020.}
    \label{fig:02_gender_over_time}
    \vspace{-0.5cm}
\end{figure}

Of 5,016 guests with gender annotations, 3,127 (62.34\%) were male and 1,880 (37.48\%) female; male appearances accounted for 65.76\% of appearances with known gender.
Nine individuals (0.18\%) carry non-binary annotations (non-binary: 4, trans woman: 2, trans man: 1, agender: 1, transmasculine: 1).
Of the 4,861 episodes with guest data, we could not identify a female guest in 16.97\%, compared to 3.64\% for male guests.
\secref{fig:02_gender_over_time} shows that the gender gap has been narrowing since 2006 but the consistent decline appears to have ceased around 2020, partly because \textit{Markus Lanz} has again featured more male guests.
A Spearman rank correlation ($\rho = -0.72$, $p < 0.001$) confirms the long-term decline; a monthly split search across all 4,819 dated episodes identifies April~2020 as the earliest month at which the post-period trend is no longer statistically significant (post-Spearman $\rho = -0.17$, $p = 0.14$; Mann-Whitney pre/post: 68.8\% vs.\ 61.9\%, $p < 0.001$), consistent with the annual-level tipping point at 2021 (post-2021 Spearman $\rho = +0.49$, $p = 0.33$).
While scalable topic analysis was deferred, early manual analysis showed that one episode with the topic \textit{Was Frauen wollen} (what women want) contained no female guests.

A methodological ambiguity warrants attention: the identified gender gap may partly reflect Wikidata's systematic underrepresentation of women\footnote{\url{https://meta.wikimedia.org/wiki/Women_in_Wikipedia/Home}} rather than editorial selection alone, since guests without Wikidata entries contribute no property data.
Yet, even if none of the 3,271 guests without gender annotations were male, the overall male appearance share would not fall under 50\%.
The true unique-guest male rate lies within $[37.7\%, 77.2\%]$, and the appearance-level rate within $[50.8\%, 73.6\%]$.
The observed rates of 62.3\% (unique guests) and 65.8\% (appearances) apply only to the Wikidata-annotated subset.
Inferring gender from ZDF and fernsehserien.de descriptions was explored but rejected: The German position and occupation labels often encode the feminine form but have no equivalent signal for masculine, introducing a directional skew rather than correcting one.
First-name-to-gender mapping was similarly ruled out.

\begin{figure}[bth]
    \vspace{-0.5cm}
    \centering
     \begin{subfigure}[b]{0.46\textwidth}
         \centering
         \includegraphics[width=\textwidth]{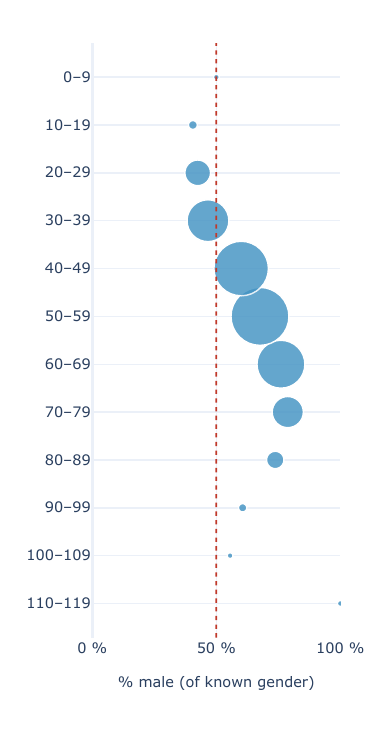}
         \caption{Male share by guest age.}
         \label{fig:03b_age_en}
     \end{subfigure}
     \hfill
     \begin{subfigure}[b]{0.53\textwidth}
         \centering
         \includegraphics[width=\textwidth]{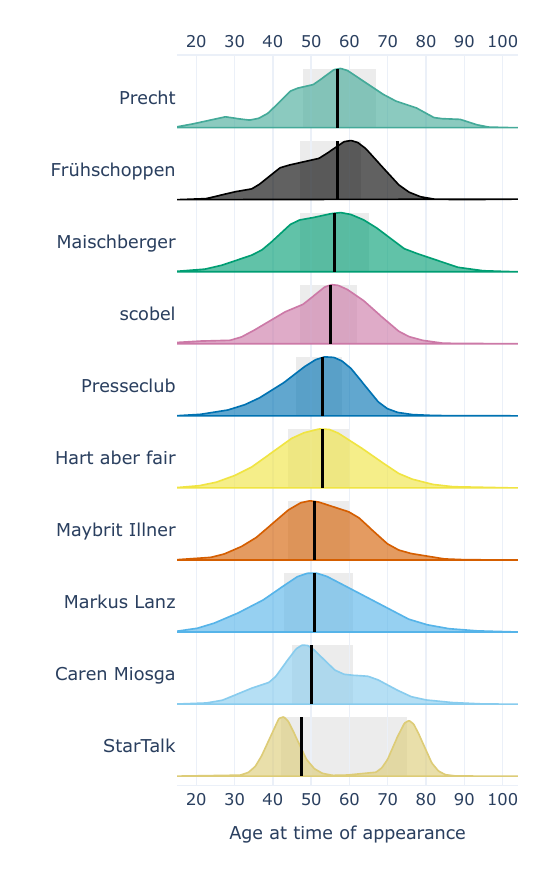}
         \caption{Age distribution per program.}
         \label{fig:01_age_ridge_plot}
     \end{subfigure}
    \caption{Age and gender distributions across the analyzed broadcasting programs.}
    \label{fig:age_and_gender}
    \vspace{-0.5cm}
\end{figure}

A distributional difference emerges in the age data (\secref{fig:age_and_gender}): male guests have a median age of 54 versus 49 for female guests, and the pattern of older men and younger women is visible across most programs.
Both a Mann-Whitney U test ($U = 44{,}154{,}210$, $p = 2.8 \times 10^{-187}$, rank-biserial $r = 0.27$, meaning male guests were older in approximately 63\% of our recorded gender-tagged appearance pairs) and a Welch's t-test on means (male: $\bar{x} = 54.8$~yrs, female: $\bar{x} = 48.8$~yrs; $t = 30.0$, $p = 2.3 \times 10^{-190}$, Cohen's $d = 0.48$) confirm this difference, a small-to-medium effect.
Science communication formats such as \textit{scobel} and particularly \textit{StarTalk} seem to bring in older expert guests;
political talk shows such as \textit{Hart aber fair} and particularly \textit{Maischberger} show the broadest age distributions, reflecting the different expertise profiles each format draws on.

\subsubsection{Occupation, Party, and Property Coverage.}

Only 55 appearances were by guests classified as scientists (Q901) directly, but 725 by economists (Q188094, a subclass), illustrating why subclass-aware queries are essential.
Yet, Wikidata's occupation taxonomy contains loops (scientist $\xrightarrow{P279}$ researcher $\xrightarrow{P279}$ academic professional $\xrightarrow{P279}$ academic $\xrightarrow{P279}$ scientist) and requires pipeline-level resolution. 
Person-to-organization relationships proved sparser than expected: \textit{Die Welt} appears only 110 times as employer despite Robin Alexander appearing 131 times and having been \textit{Die Welt}'s deputy editor-in-chief from 2019 to 2025, but with an empty Wikidata employer field.
Wikidata nonetheless proved a very benifical resource for visualization: \textit{short labels (P1813)} reduced ``Bündnis Sahra Wagenknecht -- Vernunft und Gerechtigkeit'' to ``BSW'', and party colors (\textit{P462}/\textit{P6364}/\textit{P465}) enabled consistent political-spectrum ordering across visualizations. 
Both properties would have required extensive manual curation to replicate and maintain consistency.
Additionally, labels of each visualization can be translated to any Wikidata-covered language almost automatically.
Additional statistics and visualization permutations are available in the repository\footnote{\url{https://github.com/borgnetzwerk/speaker-mining}}.

\begin{figure}[bth]
    \centering
    \caption{Wikidata property coverage for guests ($n=8{,}287$), total and per show.}
    \vspace{-0.4cm}
    \includegraphics[width=1\linewidth]{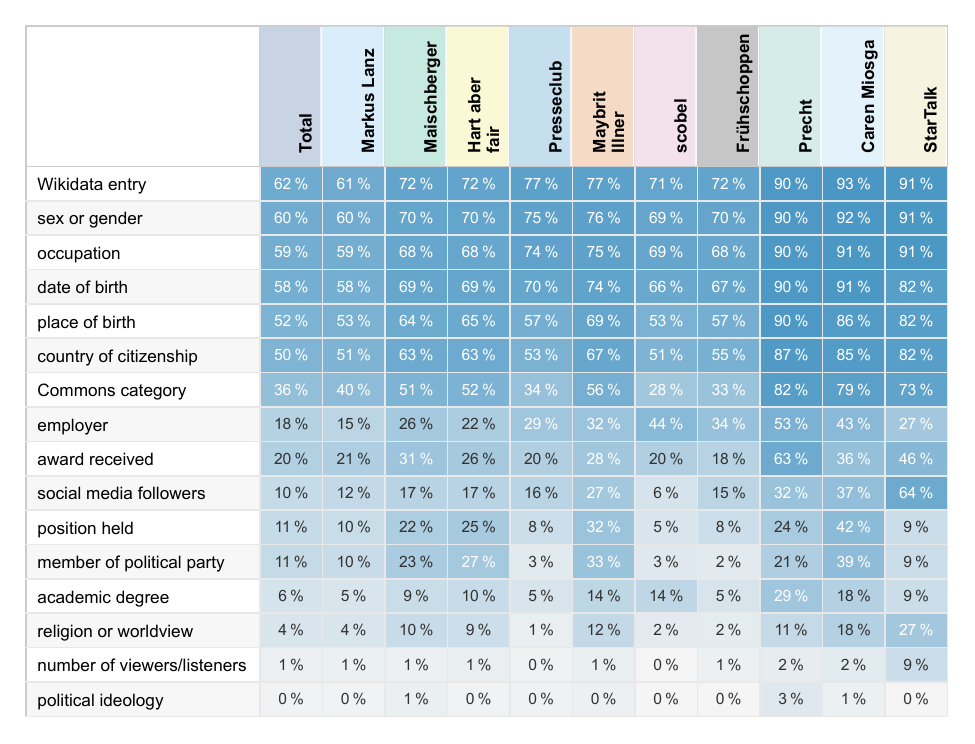}
    \label{fig:property_coverage}
    \vspace{-0.6cm}
\end{figure}

\secref{fig:property_coverage} shows Wikidata property coverage across the 10 programs for all 8,287 analyzed guests.
Date of birth (P569) is well covered for prominent guests but sparse for lesser-known individuals. 
Occupation (P106) is moderately covered, though the hierarchy issues documented above complicate aggregation. 
Coverage patterns generally align with show type: political shows have better party affiliation and position held coverage.
Interestingly, though, there seems to be no clear indicator for the small sample of scientific programs.
Coverage is computed over all guests per show; the 3,179 Tier~3 and Tier~4 guests (37.7\%) carry no Wikidata data and contribute zero coverage for all properties, pulling per-show rates below those of the Wikidata-linked subset alone.
The Wikidata graph also supports page-rank and co-occurrence analyses; structurally central entities (ZDF, Markus Lanz) rank highest, confirming the graph's integrity.

\subsection{Threats to Validity}

\textbf{Construct validity.~}
Guest appearance counting potentially conflates frequency with importance.
Wikidata properties are crowd-maintained and variable in completeness; 37.7\% of deduplicated persons are without Wikidata links and carry no property data.
All Wikidata-dependent findings (gender, age, occupation, party) are conditional on this linked subset and should not be interpreted as representative of the full guest population.
This selection bias is structural, not incidental: women, private citizens, and recent or regional public figures are systematically underrepresented in Wikidata, meaning that representational patterns observed in the linked subset might not reflect the full guest population.
Any interpretation of gender, age, or occupation distributions must account for this compounding coverage effect.
At the same time, every analysis and inspection of these Wikidata claims identifies issues and increases their overall quality.
Wikipedia has shown that crowdsourcing this knowledge leads to higher-quality knowledge representation overall. 
Our findings support this. 

\textbf{Internal validity.~}
Episodes without guests in the ZDF archive and fernsehserien.de remain unaccounted for.
Relevance propagation in Wikidata cannot reach guests unless their episode correctly references both the seed and the guest.
100 \% deduplication rate is likely neither achievable nor desirable in practice: Limiting factors include ambiguous or absent context, inaccurate source data, and most importantly GDPR considerations.
While our analysis excludes properties of individuals not captured in Wikidata, it also excludes the uncertain noise of duplicate guests and malformed textual data interpretation.
The semi-automated reconciliation was laborious, but also intentionally precision-tuned to ensure the resulting mapping is of high quality.

\textbf{External validity.~}
The pipeline was developed and evaluated primarily on the Markus Lanz corpus.
The ZDF Archive PDF structure, the Wikidata entity landscape and especially fernsehserien.de are particular for this dataset on German public television; 
but applying the pipeline to a different language context, broadcaster, or show format would only require retuning extraction regexes and seed QIDs.
The degree of manual effort required for Step~3.1.2 Reconciliation will vary substantially with corpus size and source quality.
Yet we found that this work is heavily front-loaded, and picking up where we left off should be much easier than starting from scratch, as ours work had to.

\textbf{Conclusion validity.~}
All results should be treated as indicative rather than exhaustive.
In particular, qualifier-based assessments, such as temporal properties, are implemented but not exhaustively tested, and are currently dropping potentially relevant claims from the past that qualified guests for their roles (e.g., former presidents).
The current pipeline state represents a proof of concept with quantified gaps, not a production system with validated end-to-end coverage.

\section{Discussion}\label{sec:discussion_futurework}

\textbf{FAIR data infrastructure is the bottleneck.}
A substantial share of the overall project time was spent processing raw, unstructured data sources. 
Every source had systematic gaps: the ZDF Archive had encoding failures and malformed episodes; 
fernsehserien.de had episodes without guest data;
At the same time, approaches like ours could benefit all sources to provide crowdsourced cross-verification for individual dataset improvement.
Ultimately, the root cause is structural: as stk noted~\cite{stk}, \textit{``it would make much more sense if the public broadcasters maintained this information as a knowledge graph on their own platforms and then linked the individuals via Wikidata or GND ID''}.
On the source level, we recommend conceptually merging Phase~1 and 2 into a single step that aggregates all sources and aligns them on a shared Wikibase before querying Wikidata.
Infrastructure efforts complementary to this are \href{https://blog.rufus-portal.de/}{rufus}~\cite{Blume.2024} for institutional archive access and SciCom Wiki~\cite{Wittenborg.2025} for crowdsourced curation.
Collaborative ecosystems built around linked open data already enable federated SPARQL queries linking our \href{https://speakermining.wikibase.cloud/}{speakermining.wikibase.cloud} to the Wikiverse.

At the data level, future work should focus on connecting additional sources (e.g., YouTube, Spotify) and untapped properties (e.g., topics, audience ratings), particularly on details such as qualifiers and references.
Preliminary YouTube-based transcript retrieval has already been explored for \textit{Lanz und Precht} as a proof of concept for extending the pipeline beyond broadcaster archives.
While expanding the data model accordingly, a mapping to the academic wheel of privilege~\cite{elsherifBridgingNeurodiversityOpen2022} could serve as a powerful framework for multidimensional representation analysis.
Conclusively, we reiterate our advocacy for the extensive annotation of these properties in a FAIR and GDPR-compliant manner.
Particularly on the transcript level, as well as actual transcript-based speaker (share) annotation, are promising, yet face even more complex GDPR and copyright challenges.

\textbf{Disambiguation at scale is likely frontloaded.}
The manual curation bottleneck documented in \nameref{sec:results}, and the Step~3.1.1 Alignment automation ceiling for persons, are the clearest signals that entity disambiguation cannot yet be called scalable.
At the same time, as the system improved and lessons were learned, we could reduce the number of manual matches required.
If we scale all three pillars, from data providing from broadcasters to data processing from pipelines such as ours to data curation through crowdsourced infrastructures, we assume the largely front-loaded curation task will eventually become achievable.
The Wikiverse has repeatedly shown that previously unimaginable knowledge work can be achieved when the community is provided with sufficient infrastructure.

Despite current gaps, the curated dataset already enables questions previously unanswerable at scale:
Which shows have the least gender-balanced guest profiles? 
Which occupational groups are systematically absent from political discourse?
How has guest diversity evolved over 20 years?
These are the questions motivating the SciCom KI, and they require a structured, community-driven infrastructure that this work contributes to.

\textbf{Sustainability focused via SciCom KI.}
The long-term curation for this data can be facilitated in the knowledge infrastructure for science communication.
Particularly the SciCom Wiki~\cite{WissKommProposal.2026} is designed as a sustainable, crowdsourced successor to isolated efforts like our related work: using a shared knowledge base, editorial governance, and community tooling for individual researchers and projects.
Speaker Mining provides a proof of concept for SciCom Wiki, demonstrating which data can be curated at scale, which disambiguation workflows are reusable, and which legal and technical open questions must be resolved before such a platform can operate at full scope.
The \href{https://speakermining.wikibase.cloud/}{speakermining.wikibase.cloud} instance and the pipeline developed here will be migrated into SciCom Wiki as a founding dataset, alongside the lessons learned on event-sourced extraction and the extent of Wikidata coverage into the next iteration.
These challenges require coordination with data protection expertise as well as ethics and open science standards.
Every such contribution clears the technological and legal shroud currently hindering the SciCom KI, sustainably reduces onboarding costs, and advances towards FAIR question answering on public broadcasts at scale.

\section{Conclusion}\label{sec:conclusion}

This paper presents a scaling-oriented framework for FAIR curation of public-broadcast guest data, evaluated across 15 configured broadcasting programs. 
Starting from three heterogeneous sources, we produced a unified, queryable knowledge base of 8,436 canonical persons with 23,527 appearances across 6,469 aligned episodes, published at \href{https://speakermining.wikibase.cloud/}{speakermining.wikibase.cloud}.
Our contributions are (1)~a reusable, open-source pipeline for multi-source broadcast metadata curation; (2)~a curated FAIR dataset covering 12 broadcasting programs with episode data (10 with guest data); (3)~quantitative property analyses of primarily gender and distributions across the 10 analyzed programs.
Within the Wikidata-annotated subset, a gender gap is visible in our data slice of public broadcasting: male guests account for 65\% of appearances with known gender, and a distributional age difference is observed towards older male guests.
The long-term trend toward gender parity appears to be slowing since 2021, though the recency of this period limits how firmly this can be characterized.
Our main contribution, the pipeline and infrastructure enabling this analysis, is a promising proof of concept, with much potential to grow and contribute towards a crowdsourced SciCom KI.
Complementary efforts from \href{https://blog.rufus-portal.de/}{rufus} to SciCom Wiki~\cite{Wittenborg.2025,WissKommProposal.2026} face similar technical and legal challenges ahead.
To overcome them, we must share not only data, but also open-source software, reconciliation workflows, and lessons learned.
Following the Wikiverse example, collaboration of public broadcasters and research institutions, as well as individuals across fields, must be fostered towards scaling FAIR data on public broadcasts for reliable question answering.

\begin{credits}
\subsubsection{\ackname} 
This study was funded as part of the WissKomm Wiki project, part of the Lower Saxony Research Data Management Initiative (FDM-NDS), a funding program by the Lower Saxony Ministry of Science and Culture (MWK) and the Volkswagen Foundation.
We want to acknowledge help from the ZDF archive (Veit Scheller),
Armin Müller-von Fischer (arrrrrmin) and Stefan ``stk'' Kaufmann, as well as Der Spiegel for providing data and support, and our students Jamal Eldemashki, Abdul Aahad Qureshi, and Rownak Deka, who put a lot of effort into the wikibase and dataset curation.

\subsubsection*{Use of AI tools declaration.}
During the preparation of this work, the author(s) used \textbf{GitHub Copilot}, \textbf{Claude Code}, \textbf{DeepL}, \textbf{Grammarly}, \textbf{LanguageTool} in order to: \textbf{translate text}, \textbf{grammar and spelling check}, \textbf{paraphrase and reword}, \textbf{peer review simulation}, according to the CEUR GenAI Usage Taxonomy\footnote{\url{https://ceur-ws.org/GenAI/Taxonomy.html}}.
After using these tools/services, the authors reviewed and edited the content as needed and take full responsibility for the content of the publication.

\subsubsection{\discintname}
The authors have no competing interests to declare that are
relevant to the content of this article.
\end{credits}
\bibliographystyle{splncs04}
\bibliography{bibliography}

\end{document}